\documentclass[twocolumn,showpacs,preprintnumbers,amsmath,amssymb,superscriptaddress]{revtex4}

\usepackage{graphicx}
\usepackage{dcolumn}
\usepackage{bm}

\newcommand{\bfk}{\mathbf{k}}
\newcommand{\kf}{\mathbf{k}_F}
\newcommand{\wroot}{\omega_{\bfk}^*}
\newcommand{\wkf}{\omega_{\kf}^*}
\newcommand{\qp}{\textbf{\textit{qp}} }
\newcommand{\qps}{\textbf{\textit{qp}}s }
\newcommand{\eps}{\epsilon}

\begin{document}

\title{Hidden Fermi Liquid, Scattering Rate Saturation and Nernst Effect: a DMFT Perspective}

\author{Wenhu Xu}
\author{Kristjan Haule}
\author{Gabriel Kotliar}

\affiliation{%
 Department of Physics and Astronomy, Rutgers University, 136 Frelinghuysen Rd., NJ 08854}

\date{\today}

\begin{abstract}
We investigate the transport properties of a correlated metal within dynamical mean field theory. Canonical Fermi liquid behavior emerges only below a very low temperature scale $T_{FL}$. Surprisingly the quasiparticle scattering rate follows a quadratic temperature dependence up to much higher temperatures and crosses over to saturated behavior around a temperature scale $T_{sat}$. We identify these quasiparticles as constituents of the hidden Fermi liquid. The non-Fermi liquid transport above $T_{FL}$, in particular the linear-in-$T$ resistivity, is shown to be a result of a strongly temperature dependent band dispersion. We derive simple expressions for resistivity, Hall angle, thermoelectric power and Nernst coefficient in terms of a temperature dependent renormalized band structure and the quasiparticle scattering rate. We discuss possible tests of the DMFT picture of transport using ac measurements. 
 \end{abstract}

\maketitle


Fermi liquids~\cite{landau1957theory} are good conductors. Quasiparticles (\textbf{\textit{qp}}) with a mean free path much longer than their wavelength are responsible for the electric transport, and the resistivity vanishes quadratically at low temperatures. Landau theory is very robust and when reformulated in terms of a transport kinetic equation, it can be used to describe situations where Landau \qps are strictly speaking not well defined, namely when the \qp scattering rate is comparable to their energy, such as the electron-phonon coupled system above the Debye temperature~\cite{PhysRev.134.A566}.

The metallic state of many strongly correlated materials is not described by Landau theory in a wide range of temperatures. Quadratic temperature dependence of resistivity occurs in a very narrow or vanishing range of temperatures. The interpretation of the resistivity in terms of the standard model of transport  which is based on  \qps is problematic since it  leads to mean free paths shorter than the (\textbf{\textit{qp}}) DeBroglie wavelength as stressed by Emery and Kivelson \cite{PhysRevLett.74.3253}. The transport properties of these ``bad metals'' thus requires a novel framework for their theoretical interpretation. 

Dynamical mean field theory(DMFT)~\cite{RevModPhys.68.13}, provides a non-perturbative framework for the description of strongly correlated materials. It links observable quantities to a simpler, but still interacting, reference system (a quantum impurity in a self-consistent medium) rather than to a free electron system, hence it gives access to physical regimes outside the scope of Landau theory. 

In a broad temperature range, the single-site DMFT description of the one band Hubbard model at large U and finite doping, results in transport and optical properties with anomalous temperature dependence~\cite{PhysRevB.47.3553, doi:10.1080/00018739500101526, PhysRevB.53.16214, PhysRevLett.80.4775, PhysRevB.59.1800, PhysRevB.84.035114,  PhysRevLett.110.086401, PhysRevB.87.035126}, reminiscent of those observed in bad metals. Corresponding studies of half filled metallic systems~\cite{PhysRevLett.75.105, PhysRevB.61.7996, PhysRevLett.91.016401, PhysRevLett.100.086404} also reveal bad metallic behavior in a narrower temperature region since at high temperatures the resistivity is insulating like. 

Landau \qps only emerge below an extremely low temperature, $T_{FL}$, which is much lower than the renormalized kinetic energy or Brinkman-Rice scale $T_{BR} \sim \delta W $ with $\delta$ the doping level and $W$ the bare bandwidth. $T_{BR}$ is the natural scale for the variation of physical quantities with doping at zero temperature~\cite{PhysRevB.52.17112, PhysRevB.53.16214}. A recent comprehensive  DMFT study of the Hubbard model with a semicircular bare density of states found that the transport properties above $T_{FL}$ are described in terms of resilient \qps with a strong particle-hole asymmetry~\cite{PhysRevLett.110.086401}. This asymmetry arises from the asymmetric pole structure in the self energy characterizing the proximity to the Mott insulator~\cite{PhysRevB.84.205104}. 

In this Letter we investigate the problem of bad metal transport. By expressing the DMFT transport coefficients in terms of \qp quantities we find several surprising results: a) the \qp scattering rate has a quadratic behavior for temperature much larger than $T_{FL}$ and crosses over to a saturated behavior around $T_{sat}$.
 b) The temperature dependence of the transport coefficients is anomalous (in the sense that it does not reflect the $T$ dependence of the \qp scattering rate) and 
 arises from the temperature dependent changes of the \qp dispersion near the Fermi level. 
 c) The temperature dependence of the \qp dispersion affects differently the diagonal and off-diagonal charge and thermal transport coefficients but the Mott relation~\cite{PhysRevB.21.4223, behnia2009nernst} is valid when $T_{FL}<T<T_{sat}/2$.
 
We study the one-band Hubbard Hamiltonian on the two-dimensional square lattice with nearest neighbor hopping.
\begin{equation}
 H = -t\sum_{\langle ij \rangle, \sigma}c^{\dagger}_{i\sigma}c_{j\sigma}+U\sum_{i}c^{\dagger}_{i\uparrow}c_{i\uparrow}c^{\dagger}_{i\downarrow}c_{i\downarrow}.\label{eq:hub_ham}
\end{equation}
We set the full bare bandwidth $W = 8t$ to $W=1$ as the unit of energy and temperature, and present results for $U/W = 1.75$, for which the system is a Mott insulator at half-filling. The doping level of the metallic state is fixed at $\delta = 15\%$ ($n = 0.85$). We use continuous time quantum Monte Carlo method (CTQMC)~\cite{PhysRevLett.97.076405} and the implementation of ref.~\cite{PhysRevB.75.155113} to solve the auxiliary impurity problem. We use Pad\'e approximants to analytically continue the self energy. 

\begin{figure}
 \begin{tabular}{ccc}
 \includegraphics[width=0.25\textwidth]{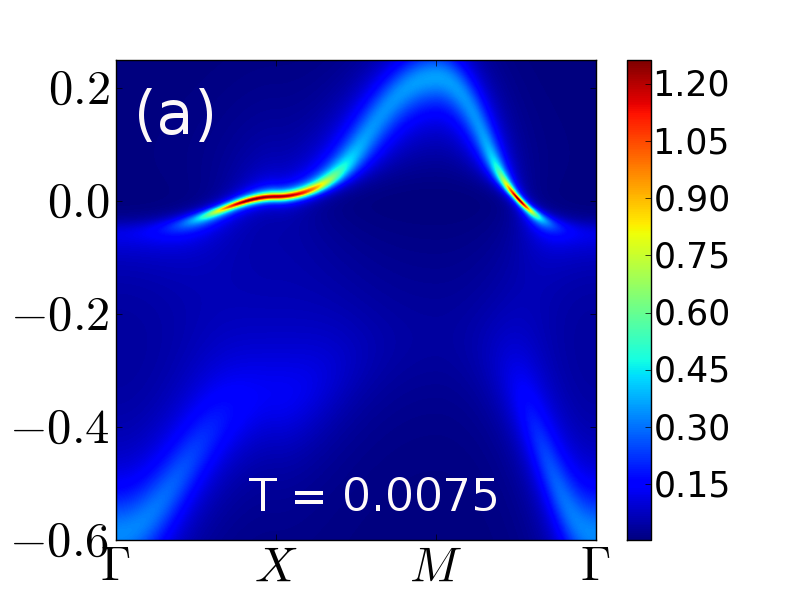} & \includegraphics[width=0.25\textwidth]{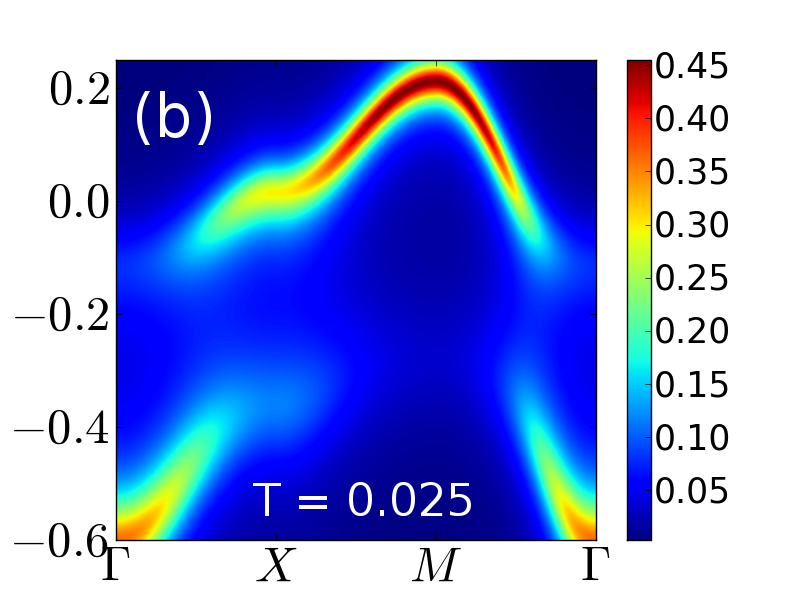} \\ \includegraphics[width=0.25\textwidth]{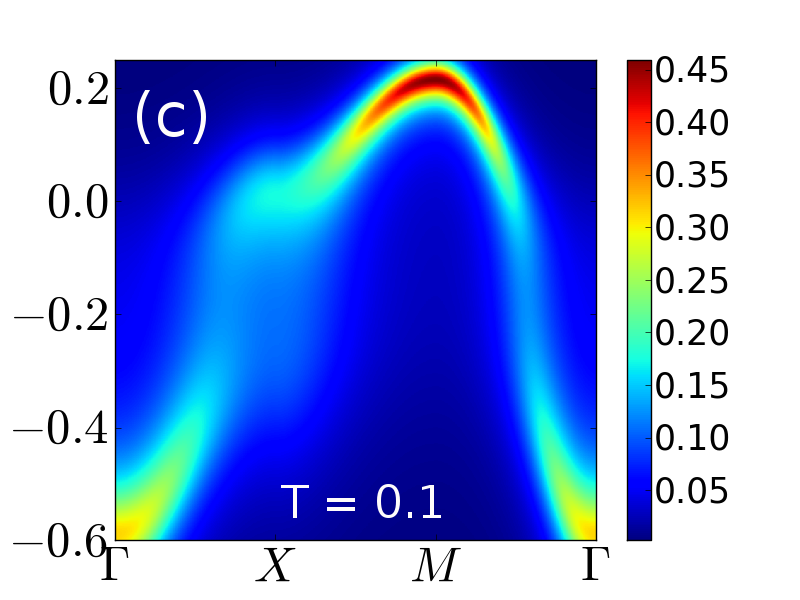} & \includegraphics[width=0.25\textwidth]{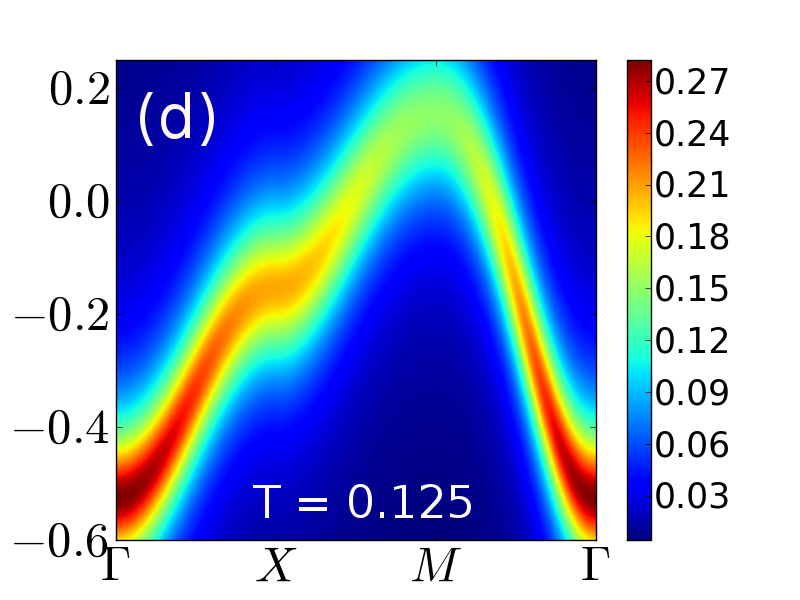} \\ 
 \includegraphics[width=0.2\textwidth]{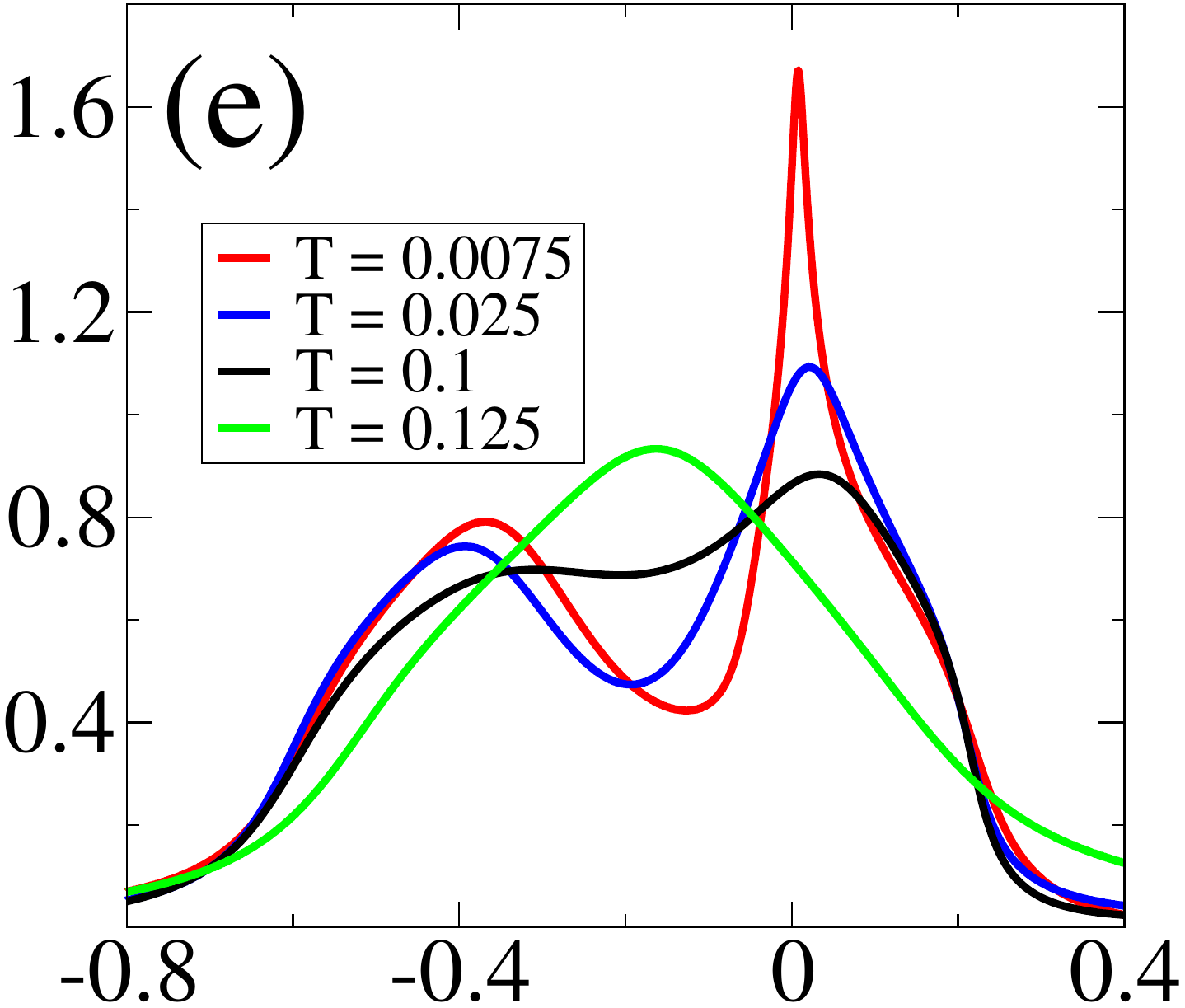} & \includegraphics[width=0.2\textwidth]{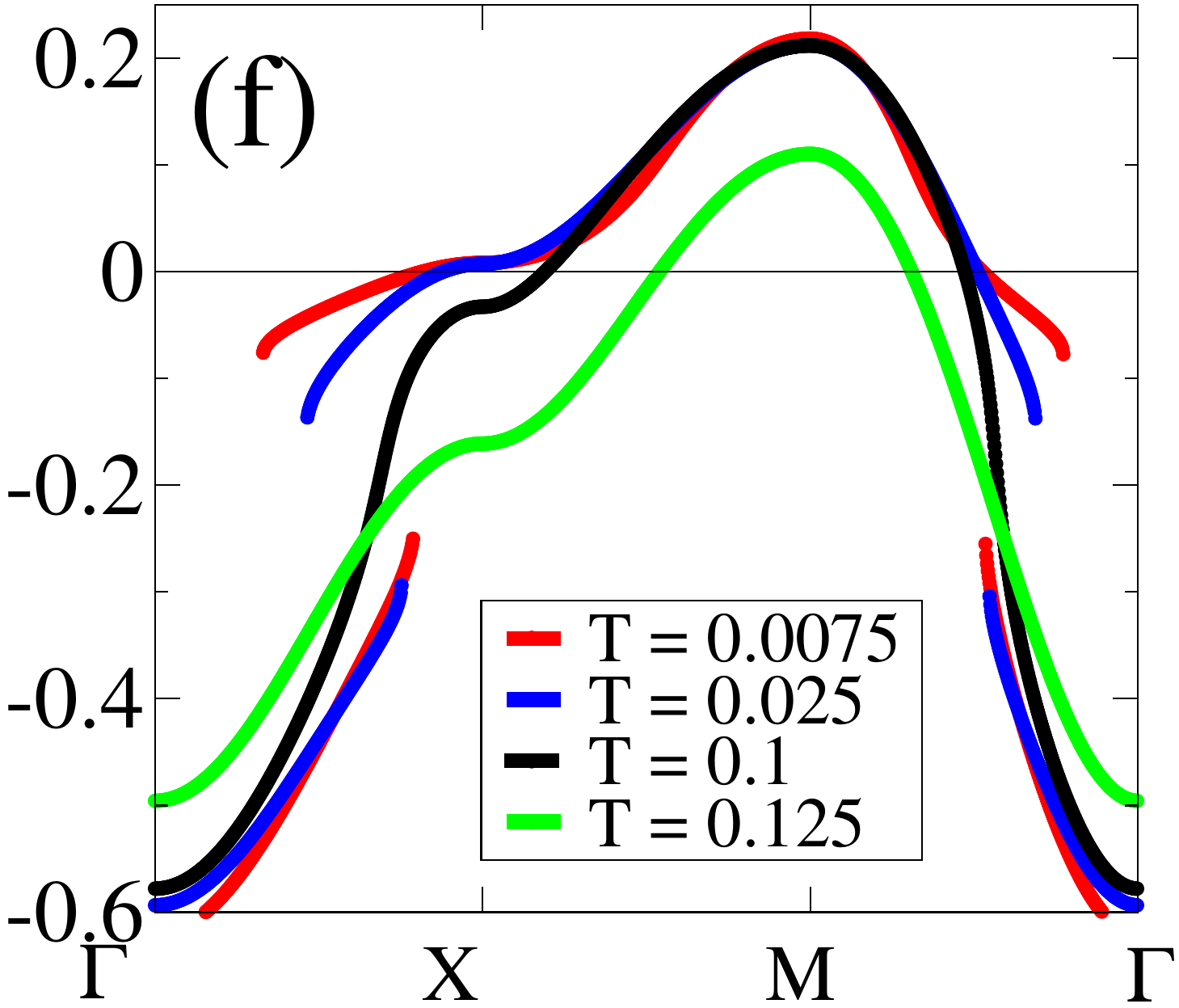} & 
 \end{tabular}
 \caption{\label{fig:qp_bands} Spectral function $A_{\bfk}(\omega)$ along $\Gamma-X-M-\Gamma$ in Brillioun zone at (a) $T = 0.0075$, (b) $T = 0.025$, (c) $T = 0.1$ and (d) $T = 0.125$. (e) Local density of states. (f) Roots of Eq.~\ref{eq:qp_root} at different temperatures.  }
\end{figure}

The one-electron spectral function is defined as 
\begin{equation}
 A_{\bfk}(\omega) = -\frac{1}{\pi}\frac{\Im\Sigma(\omega)}{(\omega+\mu-\epsilon_{\bfk}-\Re\Sigma(\omega))^2+\Im\Sigma(\omega)^2}, \label{eq:Akw}
\end{equation}
in terms of the bare band dispersion $\epsilon_{\bfk} = -\frac{1}{4}(cos(k_x)+cos(k_y))$ and the self energy $\Sigma(\omega)$. $A_{\bfk}(\omega)$ at different temperatures are plotted in Fig.~\ref{fig:qp_bands} (a-d).

Several characteristics of the evolution of $A_{\bfk}(\omega)$ with temperature are important.
The solutions of the following equation, 
\begin{equation}
 \omega+\mu(T)-\epsilon_{\bfk}-\Re\Sigma(\omega , T ) = 0, \label{eq:qp_root}
\end{equation}
faithfully reproduce the location of the peaks and how they evolve with temperature in $A_{\bfk}(\omega)$ (Fig.~\ref{fig:qp_bands}(f)). 
We do not describe in the following the upper Hubbard band, at positive energies of order $U$. 

There are two distinct temperature regimes separated by a crossover scale $T_{sat} \simeq 2T_{BR}/3 = 0.1$, 
which also sets the scale for the saturation of \qp scattering rate as will be explained later.
Above $T_{sat}$, say at $T = 0.125$, $A_{\bfk}(\omega)$ has one peak, \textit{i.e.}, Eq.~\ref{eq:qp_root} has only one root
for each $\bfk$ and displays a continuous dispersion over the whole Brillioun zone.  Below $T_{sat}$, 
Eq.~\ref{eq:qp_root} can have multiple roots. The high temperature band
breaks   into two parts, which 
together with the upper Hubbard band forms the characteristic DMFT three-peak structure of local density of states (LDOS, Fig.~\ref{fig:qp_bands}(e)). 
The breakup of these bands also leads to the separation of the optical  spectrum  into a Drude peak and a mid-infrared feature,   
characteristic of many correlated systems,  which  provided the earliest
experimental tests of the DMFT picture of correlated
materials~\cite{PhysRevB.47.3553, doi:10.1080/00018739500101526, PhysRevLett.75.105, PhysRevB.61.7996} . 

There is always a dispersive \qp feature in a $\sim k_BT$ energy window of the Fermi energy. $\omega_{\bfk}^*$ denotes  the root of Eq.~\ref{eq:qp_root} closest to the Fermi level for  a given $\bfk$. It evolves {\it continously} with temperature from zero temperature up to very high temperatures where  there is no sharp peak in the one-particle local denisty of states (LDOS, Fig.~\ref{fig:qp_bands}(e)). The character of the dispersive excitations evolves continously from strongly renormalized \qps located near the Luttinger Fermi surface with Fermi crossings around the $X$ point and on the $\Gamma-M $ line for  $T \ll T_{sat}$ 
(Fig.~\ref{fig:qp_bands}(a-b)) to holes in the lower Hubbard band (located near the $M$ point)(Fig.~\ref{fig:qp_bands}(d)) for $T \gg T_{sat}$, as the spin degrees of freedom gradually unbind from the charge, with increasing temperature. The  \qp velocity is nearly temperature independent only below $T_{FL}$ and above $T_{sat}$.  The  mass enhancement ($1/Z$), decreases with increasing temperature, from a large value $\sim 5$ below $T_{FL}$ (Fig.~\ref{fig:qp_bands}(a)) to a value $\sim 1.5 \simeq (1-n/2)^{-1}$ at high temperature (Fig.~\ref{fig:qp_bands}(d)).

We now turn to the transport properties and focus on the electric current induced by electric fields and thermal gradients. 
\begin{equation}
 \mathbf{J}_e = \bar{\sigma}^0\cdot\mathbf{E}-\bar{\sigma}^1\cdot\nabla\mathbf{T}. \label{eq:linear_response}
\end{equation}
$\bar{\sigma}^{0/1}$ is charge/thermal conductivity matrix. Several quantities of interest are resistivity($\rho$), Hall angle($\theta_H$), Seebeck coefficient($S$) and Nernst coefficient($\nu$)~\cite{behnia2009nernst}. They are representative measures of longitudinal/transverse and magneto/thermo-electric transport properties and can be expressed in terms of elements of conductivity matrices,
\begin{eqnarray}
 \rho = \frac{1}{\sigma^0_{xx}}, &\quad&\tan\theta_H = -\frac{\sigma^0_{yx}}{\sigma^0_{xx}}, \nonumber \\
 S = -\frac{\sigma^1_{xx}}{\sigma^0_{xx}}, &\quad&
 \nu = -\frac{1}{B}\left(\frac{\sigma^1_{yx}}{\sigma^0_{xx}}-\frac{\sigma^1_{xx}\sigma^0_{yx}}{(\sigma^0_{xx})^2}\right). \label{eq:transport_def}
\end{eqnarray}

Within the DMFT treatment of the one-band Hubbard model current vertex corrections vanish 
and the transport properties can be interpreted directly in terms of one-electron spectral function~\cite{PhysRevB.47.3553, PhysRevLett.80.4775, PhysRevB.59.1800},
\begin{eqnarray}
 \sigma^{\alpha}_{xx} &=& 2\pi\sum_{\bfk} \Phi^{xx}_{\bfk} \int d\omega\left(-\frac{\partial f(\omega)}{\partial \omega}\right)\left(\frac{\omega}{T}\right)^{\alpha}A^2_{\bfk}(\omega), \nonumber \\
 \frac{\sigma^{\alpha}_{yx}}{B} &=& \frac{8\pi^2}{3} \sum_{\bfk} \Phi^{yx}_{\bfk} \int d\omega\left(-\frac{\partial f(\omega)}{\partial \omega}\right)\left(\frac{\omega}{T}\right)^{\alpha}A_{\bfk}^3(\omega), \nonumber \\ \label{eq:sigma} 
\end{eqnarray}
with $\alpha = 0$ or $1$ for charge or thermal conductivity. We consider the limit of weak magnetic field, hence the off-diagonal conductivities are proportional to $B$. $\Phi^{xx}_{\bfk} = \epsilon^{x2}_{\bfk}$ and $\Phi^{yx}_{\bfk} = (\epsilon^y_{\bfk})^2\epsilon^{xx}_{\bfk}-\epsilon^y_{\bfk}\epsilon^x_{\bfk}\epsilon^{yx}_{\bfk}$ are transport functions in terms of bare band dispersion $\epsilon_{\bfk}$ and its derivatives. The derivatives are denoted by corresponding superscripts, $\epsilon^\alpha_{\bfk} = \partial\epsilon_{\bfk}/\partial k_\alpha$.

To recast Eqs.~\ref{eq:sigma} in terms of \qp, we linearize Eq.~\ref{eq:qp_root} at $\omega_{\bfk}^*$ and define $Z_{\bfk}=(1-\frac{\partial\Re\Sigma(\omega)}{\partial\omega})^{-1}|_{\omega = \wroot}$. Then the low energy part of the one-electron Green's function can be approximated as
\begin{equation}
 G_{\bfk}(\omega) \simeq \frac{Z_{\bfk}}{(\omega-\wroot)+i\Gamma^*_{\bfk}}. \label{eq:G_qp}
\end{equation}
Thus $Z_{\bfk}$ is the \qp renormalization factor (or \qp weight) and $\Gamma^*_{\bfk} = -Z_{\bfk}\Im\Sigma(\wroot)$ is the \qp scattering rate. 

Then the integrals in Eqs.~\ref{eq:sigma} can be performed analytically and lead to
\begin{eqnarray}
 \sigma^{\alpha}_{xx} &\simeq& \sum_{\bfk} \left(-\frac{\partial f(\omega)}{\partial \omega}\right)_{\omega^*_{\bfk}}\Phi_{\bfk}^{*xx}\left(\frac{\wroot}{T}\right)^{\alpha}\tau_{\bfk}^*, \nonumber\\
 \frac{\sigma^{\alpha}_{yx}}{B} &\simeq& \sum_{\bfk} \left(-\frac{\partial f(\omega)}{\partial \omega}\right)_{\omega^*_{\bfk}}\Phi_{\bfk}^{*yx}\left(\frac{\wroot}{T}\right)^{\alpha}(\tau_{\bfk}^*)^2.\label{eq:sigma_qp} 
\end{eqnarray}
$\tau^*_{\bfk} =  (\Gamma^*_{\bfk})^{-1}$ is the \qp lifetime. Transport functions are renormalized by $Z_{\bfk}$. $\Phi_{\bfk}^{*xx} = (\epsilon^{*x}_{\bfk})^2$ and $\Phi_{\bfk}^{*yx} = (\epsilon^{*y}_{\bfk})^2\epsilon^{*xx}_{\bfk}-\epsilon^{*y}_{\bfk}\epsilon^{*x}_{\bfk}\epsilon^{*yx}_{\bfk}$, with $\epsilon^{*\alpha(\beta)}_{\bfk} = Z_{\bfk}\epsilon^{\alpha(\beta)}_{\bfk}$($\alpha, \beta = x, y$). 

This reformulation leads to a transparent interpretation in terms of \qps with temperature dependent dispersion $\wroot$. Eqs.\ref {eq:sigma_qp} 
has a form similar to the solution of the kinetic equations from Boltzmann theory~\cite{ashcroft1976introduction}. The essential difference from the Prange-Kadanoff treatement of the electron-phonon problem \cite{PhysRev.134.A566} is the strong temperature dependence of the \qp dispersion brought in by $Z_{\bfk}$.

\begin{figure*}
 \begin{tabular}{cccc}
 \includegraphics[width=0.25\textwidth]{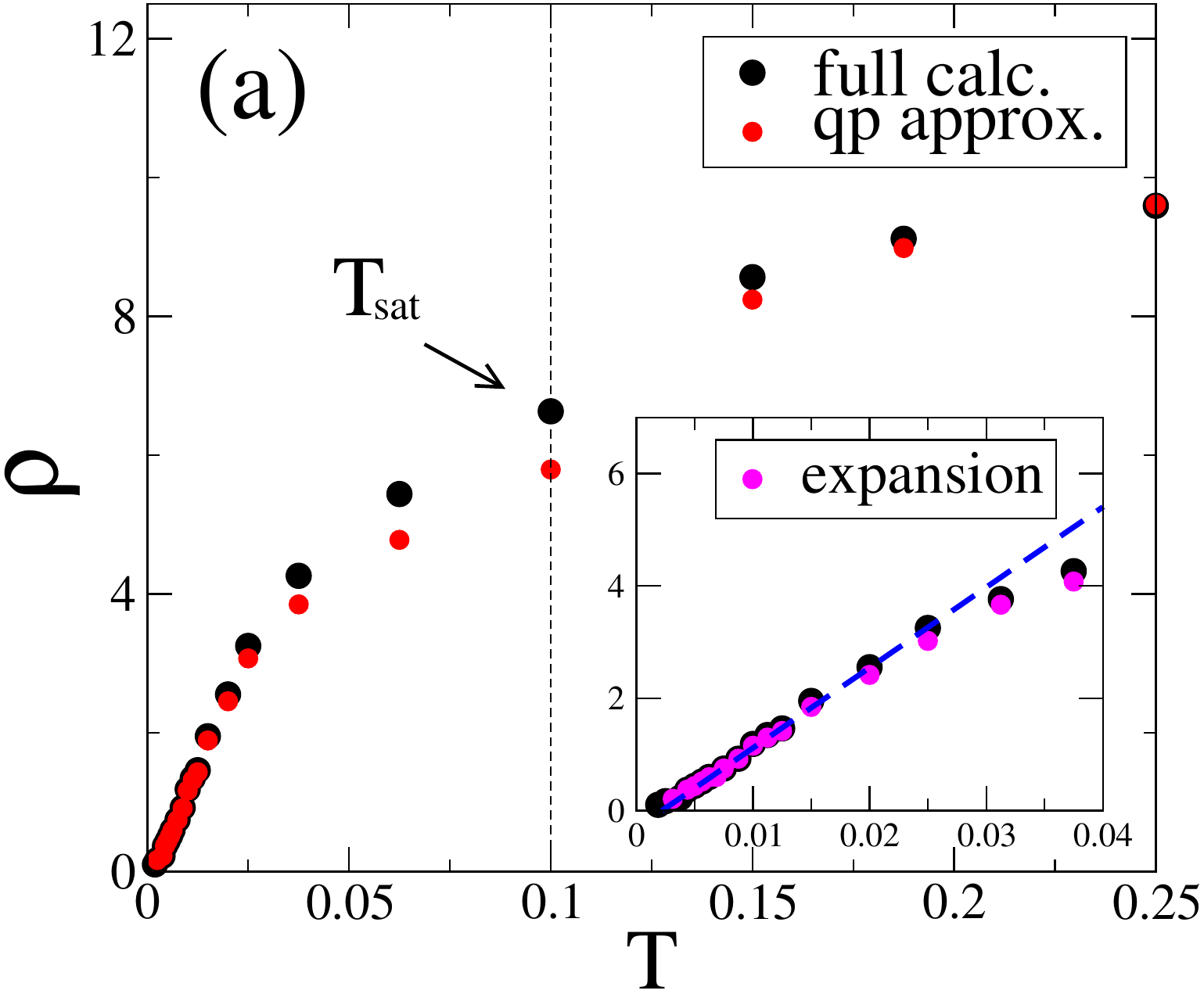} & \includegraphics[width=0.25\textwidth]{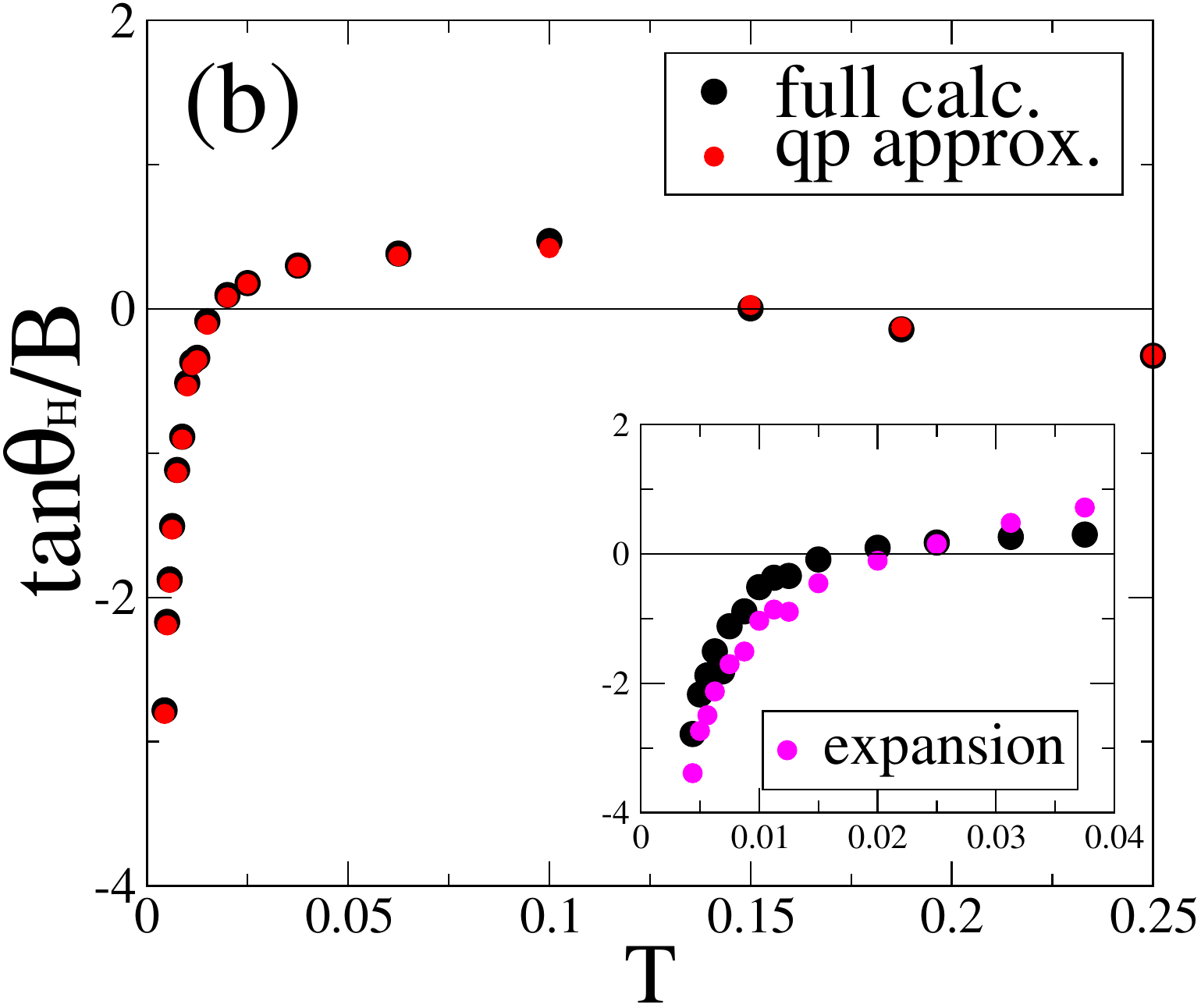} & 
 \includegraphics[width=0.25\textwidth]{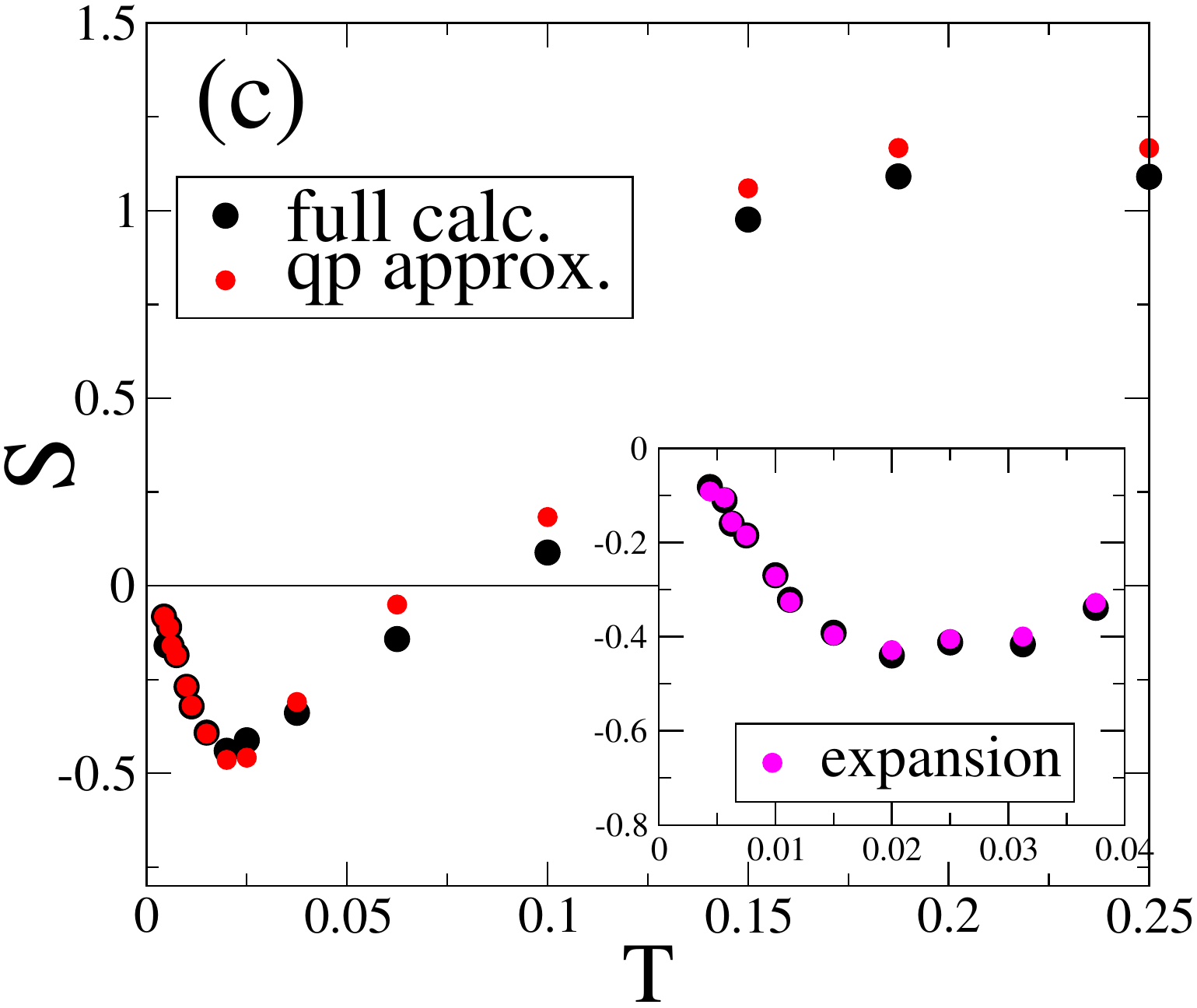} & \includegraphics[width=0.24\textwidth]{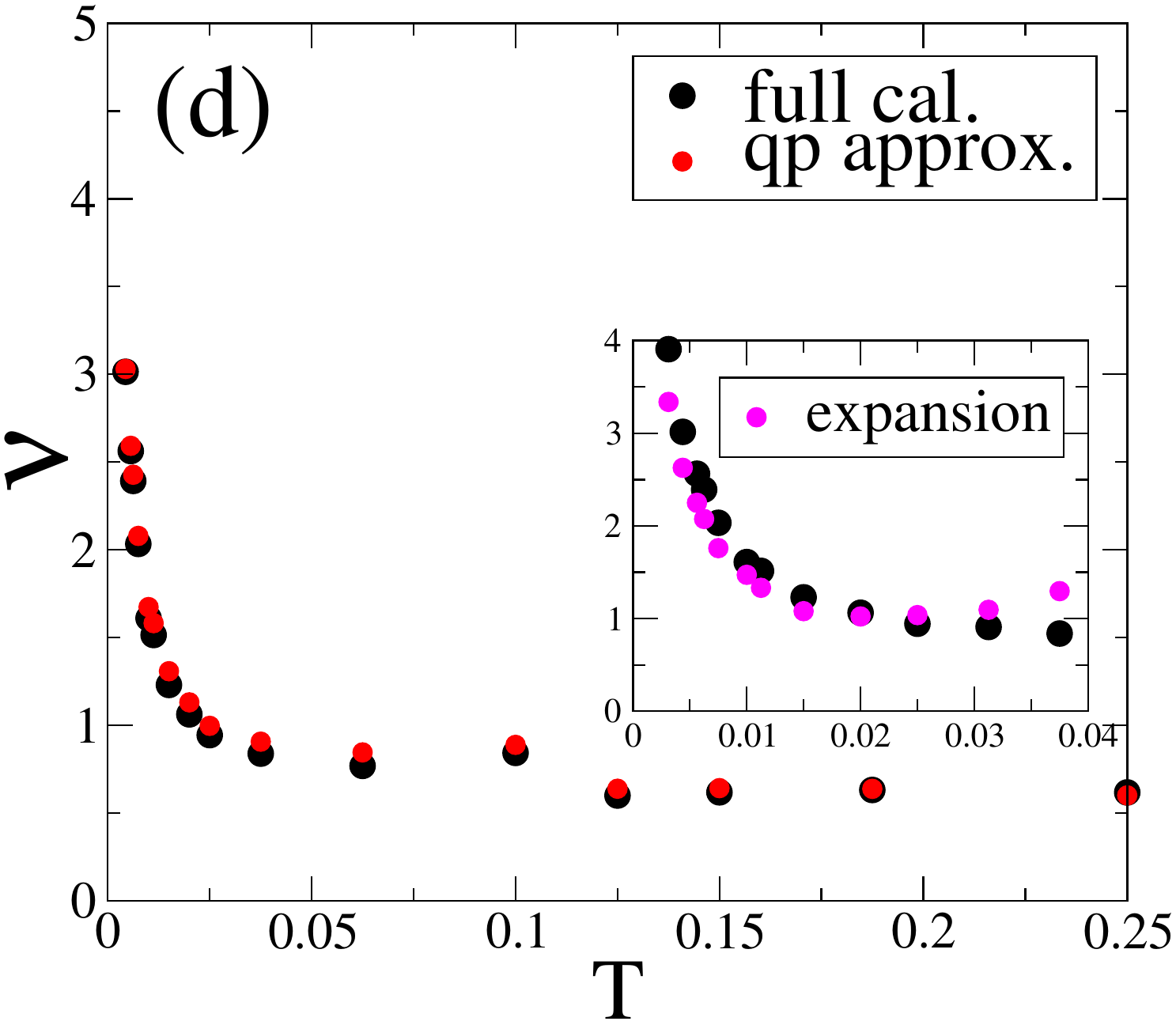}  
 \end{tabular}
 \caption{\label{fig:transport} Transport properties. (a) Resistivity. (b) Hall angle. (c) Seebeck coefficient. (d) Nernst coefficient. Black dots (``full calc.'') are obtained using Eqs.~\ref{eq:sigma}. Red dots (``qp approx.'') are obtained using Eqs.~\ref{eq:sigma_qp}. Purple dots (``expansion'') are obtained using Sommerfeld-like expansion which is detailed in Supplementary Material.}
\end{figure*}

First we validate the simplified description of transport, Eqs.~\ref{eq:sigma_qp} (``qp approx.''), by benchmarking it against the result of the exact DMFT calculation, Eqs.~\ref{eq:sigma}(``full calc.''), for the resistivity, Hall angle, Seebeck and Nernst coefficients.
The quantitative agreement between Eqs.~\ref{eq:sigma_qp} and Eqs.~\ref{eq:sigma} is evident, as shown in Fig.~\ref{fig:transport}(a)--(d). The \qp approximation faithfully reproduces the results of all transport quantities over the whole temperature range, extending to temperatures well above $T_{sat}$. 

\begin{figure*}
 \begin{tabular}{ccc}
  \includegraphics[width=0.3\textwidth]{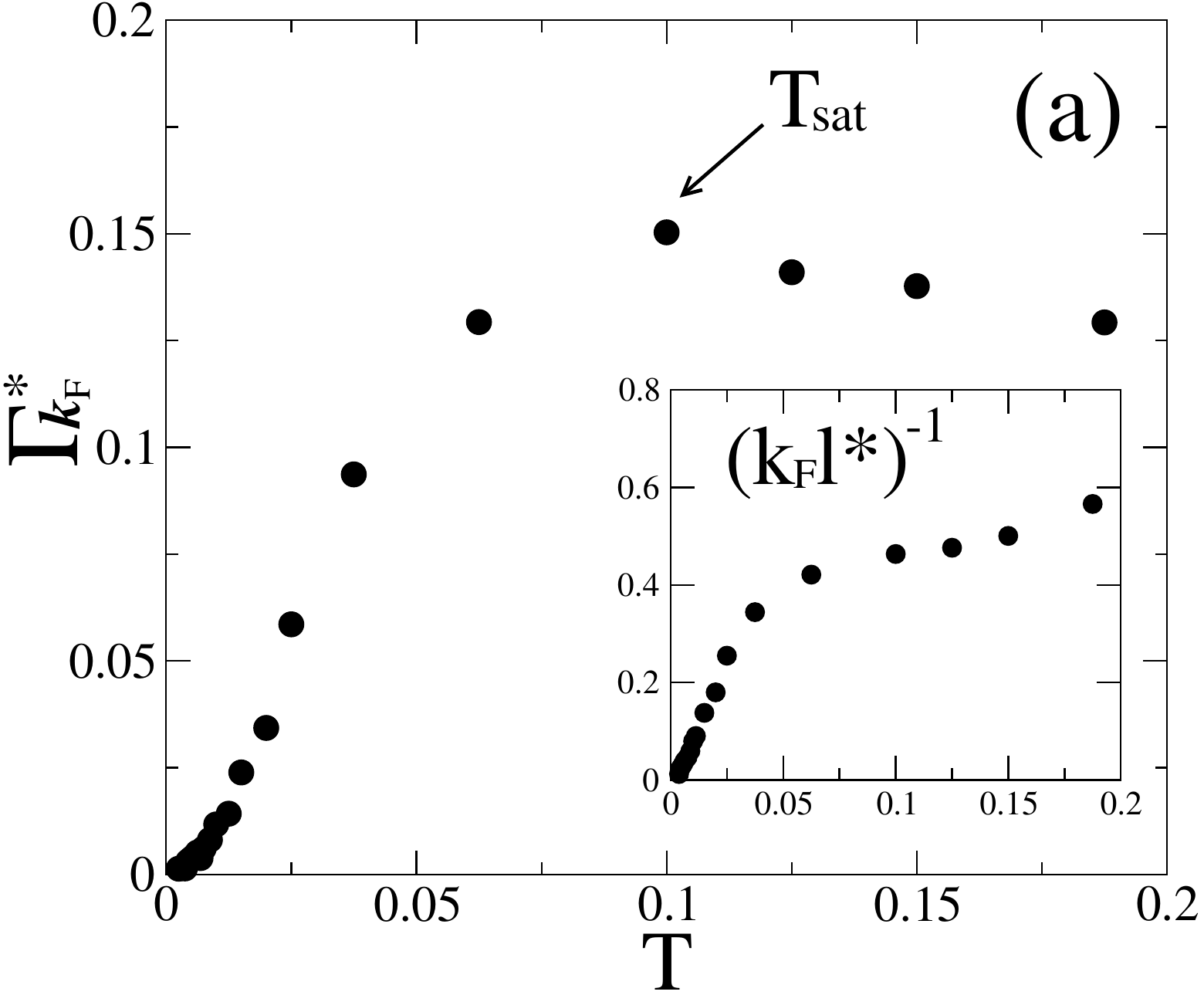} &
 \includegraphics[width=0.32\textwidth]{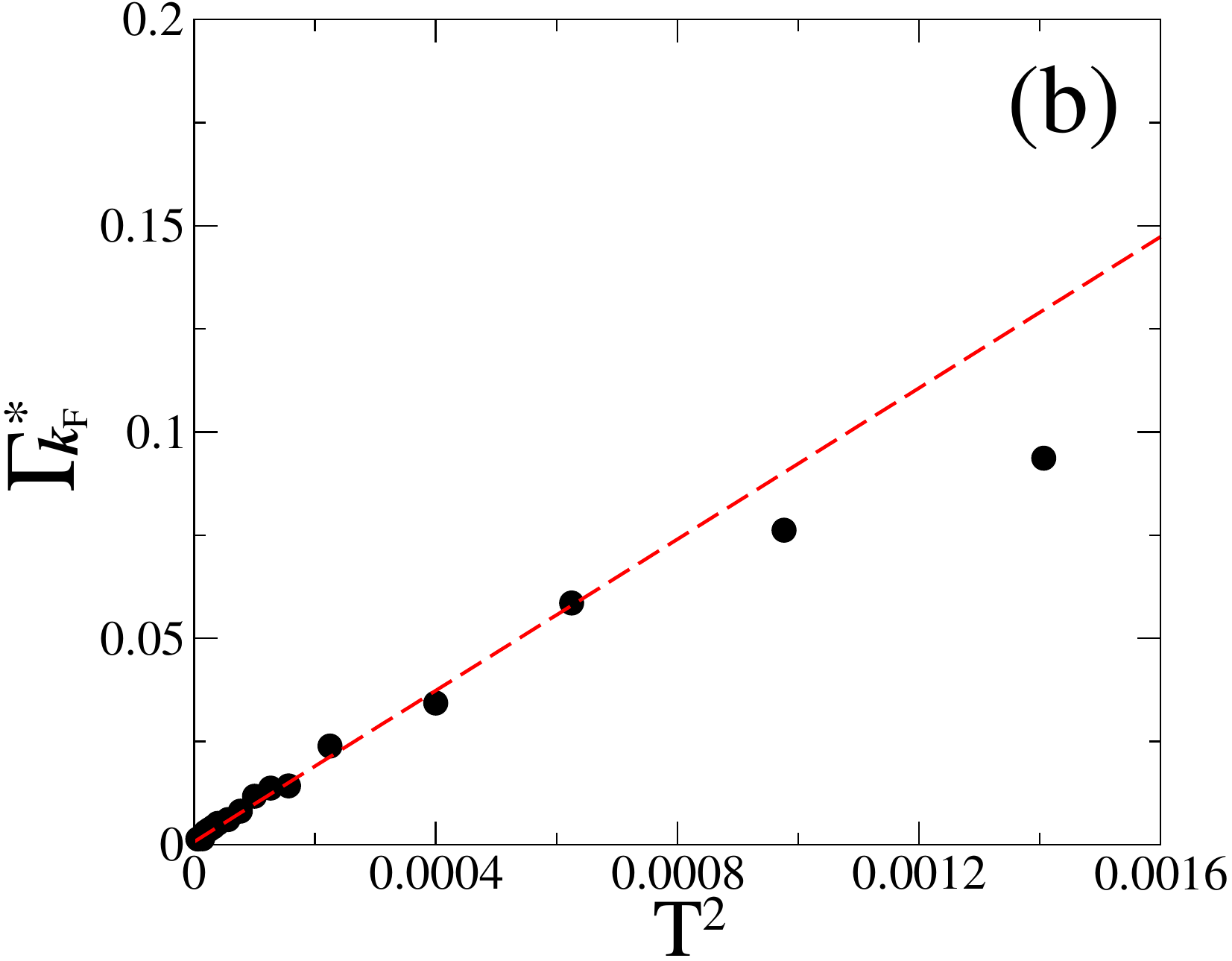} &
 \includegraphics[width=0.3\textwidth]{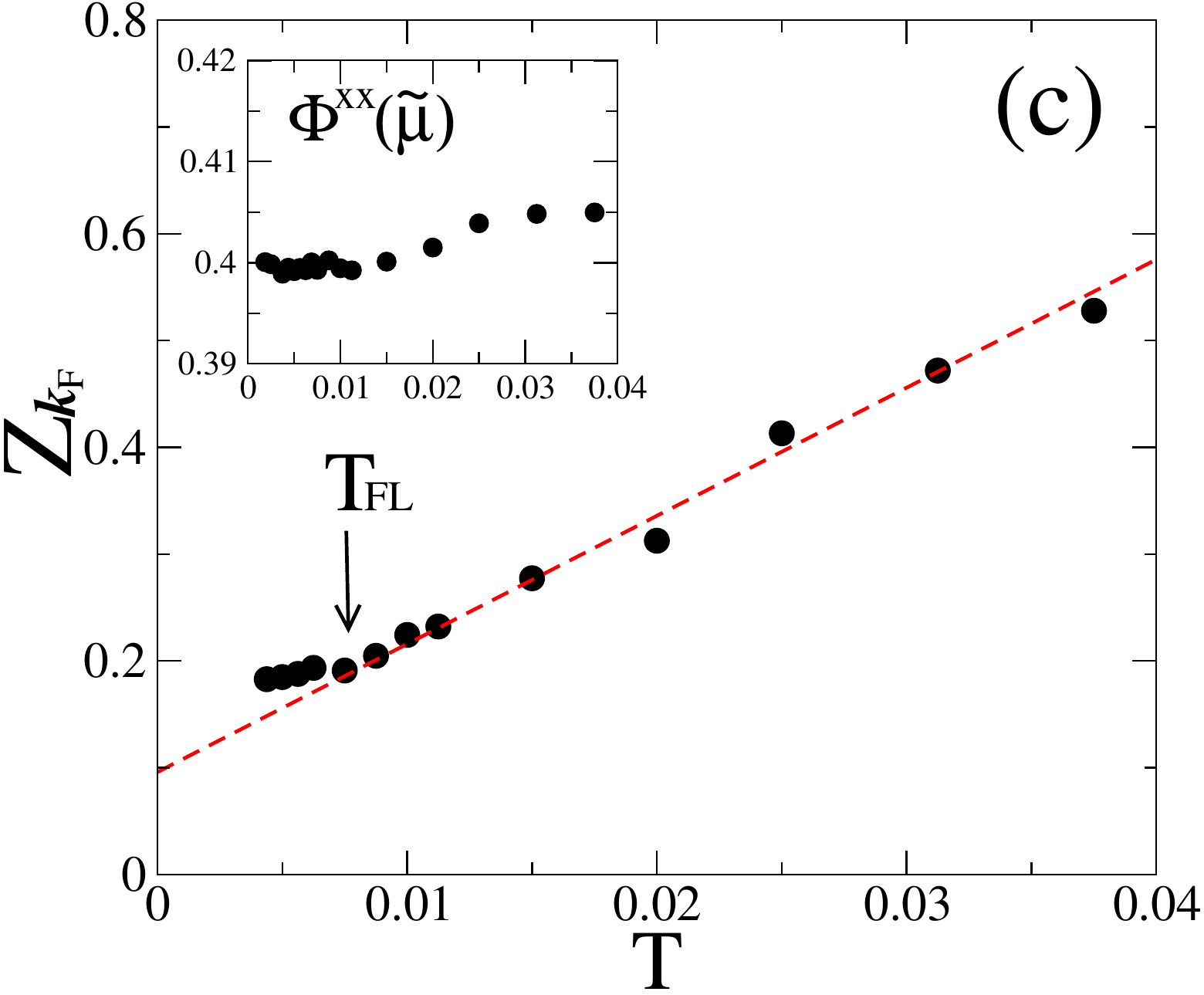} 
 \end{tabular}
 \caption{\label{fig:gamma_Z} (a) Quasiparticle scattering rate $\Gamma^*_{\kf}$. The inset shows the estimation of $(k_Fl^*)^{-1}$. (b) $\Gamma^*_{\kf}$ as a function of $T^2$ for $T \lesssim T_{sat}/2$. (c) Quasiparticle renormalization factor $Z_{\kf}$ for $T \lesssim T_{sat}/2$. }
\end{figure*}

Fig.~\ref{fig:gamma_Z}(a) shows the \qp scattering rate on Fermi surface, \textit{i.e.}, $\Gamma_{\kf}^*$ with $\wkf=0$ (For later use we also write  $\tau^*_{\kf} = (\Gamma^*_{\kf})^{-1}$ as the \qp lifetime and $Z_{\kf}$ as the renormalization factor at Fermi surface). $T_{sat}$ demarcates the non-monotonic temperature dependence of $\Gamma^*_{\kf}$. Below $T_{sat}$, $\Gamma^*_{\kf}$ increases and reaches maximum at $T_{sat}$. Above $T_{sat}$, $\Gamma^*_{\kf}$ decreases very slowly and eventually approaches to a value moderately smaller than the maximum. This confirms  that $T_{sat}$ characterizes the crossover between two distinct scattering behaviors. The inset of Fig.~\ref{fig:gamma_Z}(a) shows estimated values of $(k_Fl^*)^{-1}$ with $k_F$ an estimation of the average Fermi momentum by assuming a circular Fermi surface containing $(1-\delta)/2$ electrons and with $l^* = v^*_{\kf}\tau^*_{\kf}$ the \qp mean free path, where $v^*_{\kf} = \sqrt{\langle v^2_{\bfk}\rangle}$ with $\langle \dots\rangle$ 
averaging over the
Fermi level. At low temperatures, $(k_Fl^*)^{-1}$ increases with temperature, as expected in a good metal, and crosses over to a much slower increase, or saturated behavior around $T_{sat}/2$. Above $T_{sat}/2$, $(k_Fl^*)^{-1} \simeq 0.5$, and does not exceed the Mott-Ioffe-Regel (MIR) bound, which states that $(k_F l^*)^{-1} < 1$ in a metal. The \qps behave as expected in Boltzmann transport theory in the full temperature range, reaching the non-degenerate limit at $T \gg T_{sat}$. Notice that above $T_{FL}$, $\Im\Sigma(0)$ is not quadratic in temperature, only $\Gamma^*_{\kf} = -Z_{\kf}\Im\Sigma(0)$ is quadratic. 

The anomalies in the transport properties are the result of the strong  temperature dependence of the renormalized dispersion. This is best understood by means of a general Sommerfeld expansion of Eqs.~\ref{eq:sigma_qp}, which is explained in the Supplementary Material and works well below $T_{sat}/2$. For this purpose we define $\Phi^{*xx/yx}(\epsilon) = \sum_{\bfk} \Phi^{*xx/yx}_{\bfk}\delta(\epsilon-\wroot)$ and the energy dependent \qp lifetime $\tau^*(\epsilon) = \tau^*_{\bfk}$ when $\epsilon = \wroot$, with scattering rate $\Gamma^*(\eps) = (\tau^*(\eps))^{-1}$. For $|\eps|\lesssim T$, $\Phi^{*xx/yx}(\eps)$ is expanded to the linear order in $\eps$. To keep the asymmetry in $\Gamma^*(\eps)$ which is important for the thermoelectric transport, we expand $\Gamma^*(\eps)$ to cubic order of $\eps$, and treat the linear and cubic order as corrections to the zeroth and quadratic terms, which are dominant in the Fermi liquid regime at low temperatures. The insets in 
Fig.~\ref{fig:transport} compare the estimation using this expansion (purple dots) and the results of the full expressions (black dots). The 
agreement is evident and the expansion quantitatively captures the variation below $T_{sat}/2$.  

The inset of Fig.~\ref{fig:transport}(a) shows the linearity of resistivity, a typical non-Fermi liquid behavior~\cite{hussey2004universality}, up to $T_{sat}/4 \simeq 0.025$, as indicated by the linear fitting (blue dashed line). Surprisingly the \qp scattering rate $\Gamma^*_{\kf}$ has a quadratic temperature dependence also up to $T_{sat}/4$ (Fig.~\ref{fig:gamma_Z}(b)). This is due to the strong temperature dependence of $Z_{\kf}$ (Fig.~\ref{fig:gamma_Z}(c)). In fact, the leading order in the expansion gives
\begin{equation}
 \rho \simeq (Z_{\kf}\Phi^{xx}(\tilde{\mu})\tau^*_{\kf})^{-1}, \label{eq:sommer_rho}
\end{equation}
where $\Phi^{xx}(\tilde{\mu}) = \sum_{\bfk}\Phi^{xx}_{\bfk}\delta(\tilde{\mu}-\epsilon_{\bfk})$ with $\tilde{\mu} = \mu-\Re\Sigma(0)$ and we have used $\Phi^{*xx}(0) = Z_{\kf}\Phi^{xx}(\tilde{\mu})$. $Z_{\kf} \simeq 0.1+12T$ for $T_{FL} < T < T_{sat}/4$, leads to the quasilinear resistivity and also affects all other transport coeffecients in Eqs.~\ref{eq:sigma_qp}. The temperature dependence of $Z_{\kf}$ becomes negligible only below the Fermi liquid temperature $T_{FL} \simeq T_{sat}/15$. $\Phi^{xx}(\tilde{\mu})$ is very weakly temperature dependent as shown in the inset of Fig.~\ref{fig:gamma_Z}(c). Above $T_{sat}$, the resistivity is quasilinear in temperature with a slope smaller than that below $T_{sat}/4$, while the \qp scattering rate is saturated. The expansion cannot be used at high temperatures, but the discrepancy between the scattering rate and the resistivity, can be traced again to the temperature dependence of the dispersion, which in this regime is due to the  
 the shift of $\tilde{\mu}$ with temperature  and not to changes in \qp band structure. 

Similarly, the leading order in the Hall angle (Fig.~\ref{fig:transport}(b)) is given by $\tan\theta_H/B \simeq Z_{\kf}\Phi^{yx}(\tilde{\mu})\tau^*_{\kf}/\Phi^{xx}(\tilde{\mu})$ and indicates the sign change at $T\simeq T_{sat}/4$ is due to the sign change in $\Phi^{yx}(\tilde{\mu})$, a consequence of the evolution of Fermi surface from a hole-like one to an electron-like one. For the Seebeck coefficent (Fig.~\ref{fig:transport}(c)), the expansion leads to
\begin{equation}
 S \simeq \left(-\frac{\pi^2}{3}T\right)\left(\frac{d\ln \Phi^{*xx}(0)}{d\eps}+\frac{d \ln \tau^{*}(0)}{d\eps}\right).
\end{equation}
The asymmetry in scattering rate competes with the asymmetry in the \qp band structure, hence instead of sign change, $S$ shows non-monotonic temperature dependence below $T_{sat}/2$. 

The Nernst coefficient $\nu$ (Fig.~\ref{fig:transport}(d)) rises steeply below $T_{sat}/4$, and provides a good probe of the temperature dependence of $\tau^*_{\kf}$. The leading orders in the expansion gives
\begin{equation}
 \nu \simeq \left(-\frac{\pi^2}{3}T\right)\left[\tau^*_{\kf}\frac{d}{d\eps}\left(\frac{\Phi^{*yx}(0)}{\Phi^{*xx}(0)}\right)+\frac{\Phi^{*yx}(0)}{\Phi^{*xx}(0)}\frac{d\tau^*(0)}{d\eps}\right]. \label{eq:sommer_nernst}
\end{equation}
 In the square lattice near hall-filling, the asymmetry in band structure dominates and leads to $\nu \propto \tau^*_{\kf}T \propto 1/T$. This rise is seen in many materials~\cite{behnia2009nernst} before $\nu$ drops linearly in $T$ at very low temperature~\footnote{Disorder in real materials cuts off the divergence of $\tau^*_{\kf}$ and $\nu$ is thus linear in $T$ at very low temperatures. This linearity is sometimes taken as a signature of Fermi-liquid behavior.}.  

Further studies should be carried out to ascertain to which extent the DMFT description of transport applies to real materials, but the strong similarities between the experimental features revealed in the phenomenological picture in ref.~\cite{hussey2004universality} and our results are encouraging. 
AC transport measurements can be used to extract the temperature dependence of $\tau^*_{\kf}$. At low frequency, the optical conductivity is parametrized as~\cite{van2003quantum, RevModPhys.83.471}
$\sigma(\omega) = \frac{\omega^{*2}_{opt}}{4\pi}\left(-i\omega+\frac{1}{\tau^*_{opt}}\right)^{-1}$
, with $\omega^{*2}_{opt} \simeq 8\pi\Phi^{*xx}(0) = 8\pi Z_{\kf}\Phi^{xx}(\tilde{\mu})$ and $\tau^*_{opt} \simeq \tau^*_{\kf}/2$. Similarly in AC Hall effect~\cite{drew1999ac}, $ B\tan\theta_H(\omega) = \frac{\omega_H^{*2}}{4\pi}\left(-i\omega+\frac{1}{\tau^*_{H}}\right)^{-1}$ follows, with $\omega_H^{*2} \simeq 8 \pi\frac{\Phi^{*yx}(0)}{\Phi^{*xx}(0)} = 8\pi \frac{Z_{\kf}\Phi^{yx}(\tilde{\mu})}{\Phi^{xx}(\tilde{\mu})}$ and $\tau^*_{H} \simeq \tau^*_{\kf}/2$. Frequency dependent thermoelectric measurements would give additional information on the asymmetry of the \qp dispersion and scattering rate.

The extension from model Hamiltonians to the LDA+DMFT framework is straightforward. It can be used to separate the temperature dependence of transport coefficients arising from the temperature dependence of the \qp band and that of the scattering rate, in materials such as the ruthenates~\cite{PhysRevLett.106.096401}, the vanadates~\cite{PhysRevLett.98.166402} and the nickelates~\cite{PhysRevB.83.075125, PhysRevB.61.7996, PhysRevB.85.125137} for which the LDA+DMFT description is known to provide an accurate zeroth order picture of numerous properties~\cite{PhysRevB.61.7996}.
Recent experiments on cuprates~\cite{2012arXiv1207.6704M} have revealed evidence for temperature dependence of $\omega_{opt}^{*2}$ and a $T^2$ scattering rate over a broad range of temperatures. These materials require cluster DMFT studies to describe their momentum space differentiation. Still, it is tempting to interpret the transport properties in terms of \qps to provide an effective description of the transport. Indeed the \qp scattering rate computed in the t-J model in ref.~\cite{PhysRevB.76.104509}, exhibits the saturation behavior described in this work and it would be interesting to re-analyze the results in terms of the \qps of the hidden Fermi liquid. Our findings  are related  to two earlier theoretical proposals. Anderson introduced the idea of a hidden Fermi liquid~\cite{PhysRevB.78.174505, PhysRevLett.106.097002}, requiring $Z_{\kf}$ strictly vanishing at $T=0$ in the normal state. Alternatively, our results could be cast into the framework of the extremely correlated Fermi liquid
\cite{PhysRevLett.107.056403} by the temperature 
dependence of the caparison function.

Acknowledgement: this work was supported by NSF grant No. DMR-0906943 and No. DMR-0746395. We acknowledge useful discussions with D. N. Basov, X. Y. Deng, A. Georges, A. Kutepov, J. Mravlje and A.-M. S. Tremblay.

\bibliography{reference.bib}

\end{document}


\title{Hidden Fermi Liquid, Scattering Rate Saturation and Nernst Effect: a DMFT Perspective \\ Supplementary Material}

\author{Wenhu Xu}
\author{Kristjan Haule}
\author{Gabriel Kotliar}

\affiliation{%
 Department of Physics and Astronomy, Rutgers University, 136 Frelinghuysen Rd., NJ 08854}

\date{\today}
\maketitle

In this Supplementary Material, we consider the low temperature behavior of the integral
\begin{equation}
 I_n = \int \frac{d\eps}{T} \left(\frac{\eps}{T}\right)^n \frac{1}{4\cosh^2(\frac{\eps}{2T})}\frac{F(\eps)}{G(\eps)}. \label{sup:eq:In}
\end{equation}
$I_n$ evaluates the conductivities when $F(\eps)$ and $G(\eps)$ are chosen properly. For $F(\eps) = \Phi^{*xx}(\eps)$ and $G(\eps) = \gme = (\tau^{*}(\eps))^{-1}$, $I_0$ gives the longitudinal conductivity $\sigma_{xx}^0$ and $I_1$ gives the thermal counterpart $\sigma^1_{xx}$. For $F(\eps) = \Phi^{*yx}(\eps)$ and $G(\eps) =(\gme)^2$, $I_0$ gives transverse conductivity $\sigma^0_{yx}$ and $I_1$ gives $\sigma^1_{yx}$. 

At low temperature, $\Phi^{*xx}(\eps)$ and $\Phi^{*yx}(\eps)$ are regular functions that vary smoothly for $|\eps| \lesssim T$. Thus for the purpose of evaluating the intergral, they can be expanded as,
\begin{eqnarray}
 \phixx &\simeq& \xxzero + \xxone \eps,  \\
 \phiyx &\simeq& \yxzero + \yxone \eps,
\end{eqnarray}
where $\xxone = d\xxzero/d\eps$ and $\yxone = d\yxzero/d\eps$. 

The quasiparticle scattering rate $\gme$ in a Fermi liquid is dominated by the particle-hole symmetric part at low temperature, and has the quadratic form   
\begin{equation}
 \gme \simeq g_0 + g_2\eps^2. \label{sup:eq:gamma_FL}
\end{equation}
In a clean system, $g_0 \propto T^2$ and consequently, $\gme^{-1}$ becomes a Lorentzian function with vanishing width when $T \rightarrow 0$. Hence $\gme^{-1}$ is asymptotically singular for $|\eps|\lesssim T$ and the standard Sommerfeld treatment is no longer appropriate~\cite{PhysRevLett.80.4775}. Furthermore, the particle-hole asymmetric part in $\gme$ is in particular important for thermoelectric transport. The asymmetric part can be approximated as the linear and cubic order in $\eps$ as corrections to the quadratic scattering rate~\cite{haule2009thermoelectrics}. Then we write 
\begin{equation}
 \gme \simeq g_0 + g_2\eps^2 + g_1 \eps + g_3\eps^3. \label{sup:eq:gamma}
\end{equation}
This also leads to the corrections to the Lorentzian form of $\gme^{-1}$, which are expressed as
\begin{eqnarray}
 \frac{1}{\gme} &\simeq& \frac{1}{g_0 + g_2 \eps^2} - \frac{g_1\eps}{(g_0 + g_2 \eps^2)^2} - \frac{g_3\eps^3}{(g_0 + g_2 \eps^2)^2}, \label{sup:eq:gamma}\\
 \frac{1}{\gme^2} &\simeq& \frac{1}{(g_0 + g_2 \eps^2)^2} - \frac{2g_1\eps}{(g_0 + g_2 \eps^2)^3} - \frac{2g_3 \eps^3}{(g_0 + g_2 \eps^2)^3}. \label{sup:eq:gamma2} \nonumber \\
\end{eqnarray}

By changing the variable $\eps = xT$, we define the following integral,
\begin{equation}
 E^n_m = \int dx \frac{1}{4\cosh^2(x/2)} \frac{x^n}{(g_0/T^2 + g_2 x^2)^m}, \label{sup:eq:E_nm} 
\end{equation}
and the longitudinal conductivity is estimated by
\begin{eqnarray}
 \sigma_{xx}^0 &\simeq& \frac{1}{T^2}\left(\xxzero E^0_1 - g_1 \xxone E^2_2 \right. \nonumber \\
 {}&&\left. - g_3 T^2 \xxone E^4_2 \right).  \label{sup:eq:sigma_xx}
\end{eqnarray}
The first term is determined by the leading quadratic term in Eq.~\ref{sup:eq:gamma}. The second and third terms are determined by the linear and cubic corrections, \textit{i.e.}, proportional to $g_1$ and $g_3$ respectively. Similarly, for other conductivities, we have
\begin{eqnarray}
 \sigma_{xx}^1 &\simeq& \frac{1}{T}\left(\xxone E^2_1 - \frac{g_1}{T^2} \xxzero E^2_2 \right. \nonumber \\
 {}&& \left. -g_3 \xxzero E^4_2\right), \label{sup:eq:alpha_xx} 
\end{eqnarray}
\begin{eqnarray}
 \sigma_{yx}^0 &\simeq& \frac{1}{T^4}\left(\yxzero E^0_2 - 2g_1 \yxone E^2_3 \right. \nonumber \\
 {}&& \left. - 2g_3 T^2 \yxone E^4_3\right), \label{sup:eq:sigma_yx}
\end{eqnarray}
\begin{eqnarray}
 \sigma_{yx}^1 &\simeq& \frac{1}{T^3}\left(\yxone E^2_2 - \frac{2g_1}{T^2}\yxzero E^2_3 \right. \nonumber \\
 {}&& \left. - 2g_3 \yxzero E^4_3\right). \label{sup:eq:alpha_yx}
\end{eqnarray}
In each expansion, the second and third term will be simply denoted as the $g_1$ and $g_3$ term. 

In general, $g_0$, $g_1$, $g_2$ and $g_3$ vary with temperature, thus $E^n_m$ also depends on $T$ in a non-trivial way. But for a Fermi liquid without impurity scattering, when $T \rightarrow 0$, $(g_0 + g_2\eps^2) \propto (\pi^2 T^2 + \eps^2)$, hence $E^n_m$ is a constant independent of $T$. Besides, $g_1 \propto T^2$ as $T \rightarrow 0$~\cite{haule2009thermoelectrics}. We obtain the low-temperature leading terms of conductivities. $\sigma_{xx}^0$ and $\sigma_{yx}^0$ are dominated respectively by the $T^{-2}$ and $T^{-4}$ divergence in the first term. The prefactors are determined by the \qp band structure ($\xxzero$ and $\yxzero$). For the thermoelectric transport, all three terms lead to the same power law ($T^{-1}$ for $\sigma^{1}_{xx}$ and $T^{-3}$ for $\sigma_{yx}^1$), and the prefactors are determined by the asymmetry of \qp band structure ($\xxone$ and $\yxone$) and scattering rate ($g_1$ and $g_3$). 

The Sommerfeld expansion is equivalent to neglecting the $g_2\eps^2$ term in the expansion of scattering rate, Eq.~\ref{sup:eq:gamma} and Eq.~\ref{sup:eq:gamma2}. Above $T_{FL}$, $g_2$ is small, and the Sommerfeld expansion can give quantitative estimation that correctly captures the variation of conductivities with temperature. The Mott relations for thermoelectric transport, \textit{i.e.}, 
\begin{equation}
 I_0 \simeq \frac{F(0)}{G(0)}\quad \text{and} \quad I_1 \simeq \frac{\pi^2}{3}T\frac{d}{d\eps}\left(\frac{F(0)}{G(0)}\right),
\end{equation}
are based on the Sommerfeld expansion~\cite{PhysRevB.21.4223, behnia2009nernst}, thus they still provide convenient rules for estimating the Seebeck and Nernst coefficient when $T_{FL} < T < T_{sat}/2$. Below $T_{FL}$, the $g_2\eps^2$ term will lead to different prefactors (Eq.~\ref{sup:eq:E_nm}), but the Fermi liquid power law at low temperature is not changed. 

 \begin{figure}
  \begin{tabular}{cc}
   \includegraphics[width=0.22\textwidth]{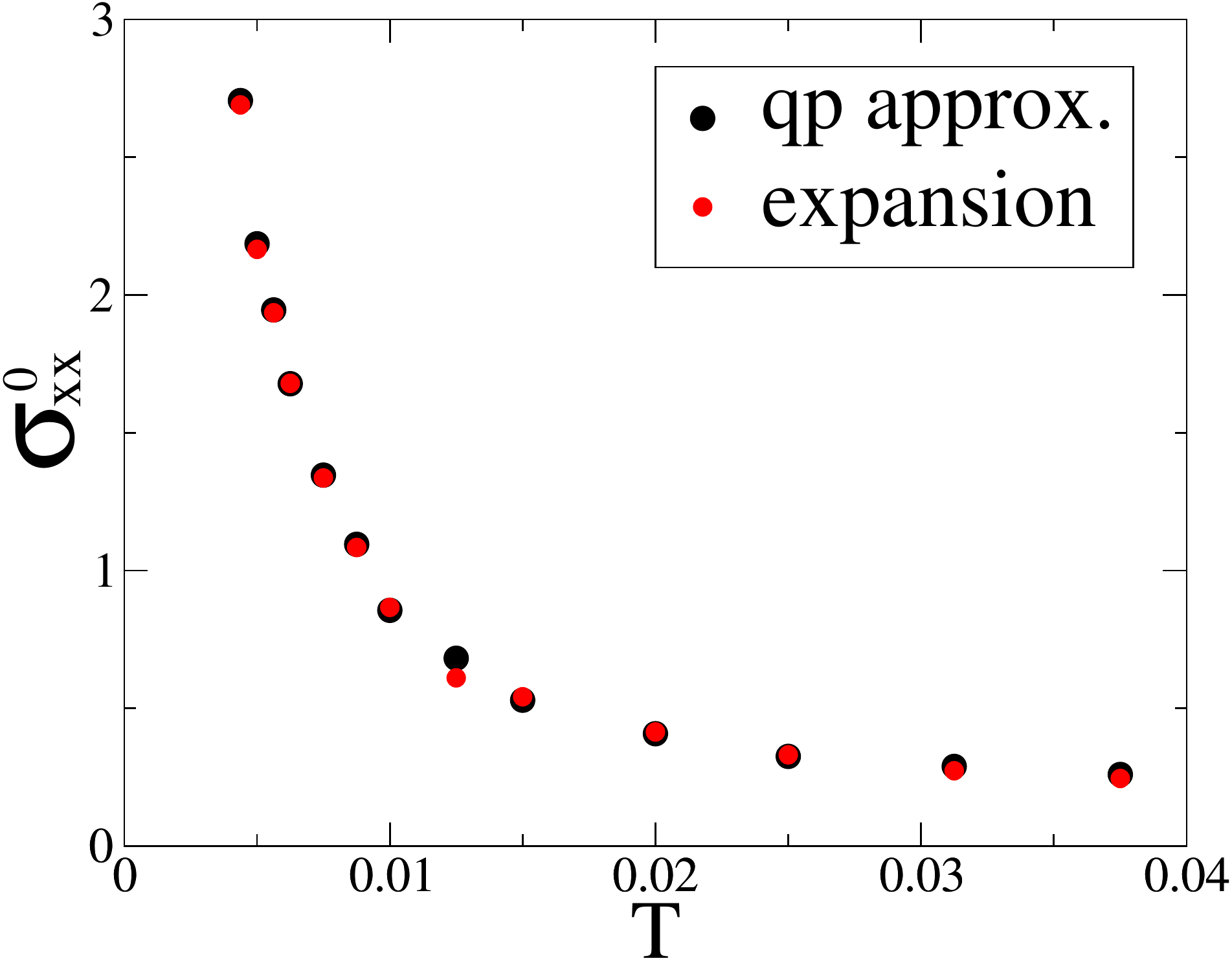} & \includegraphics[width=0.22\textwidth]{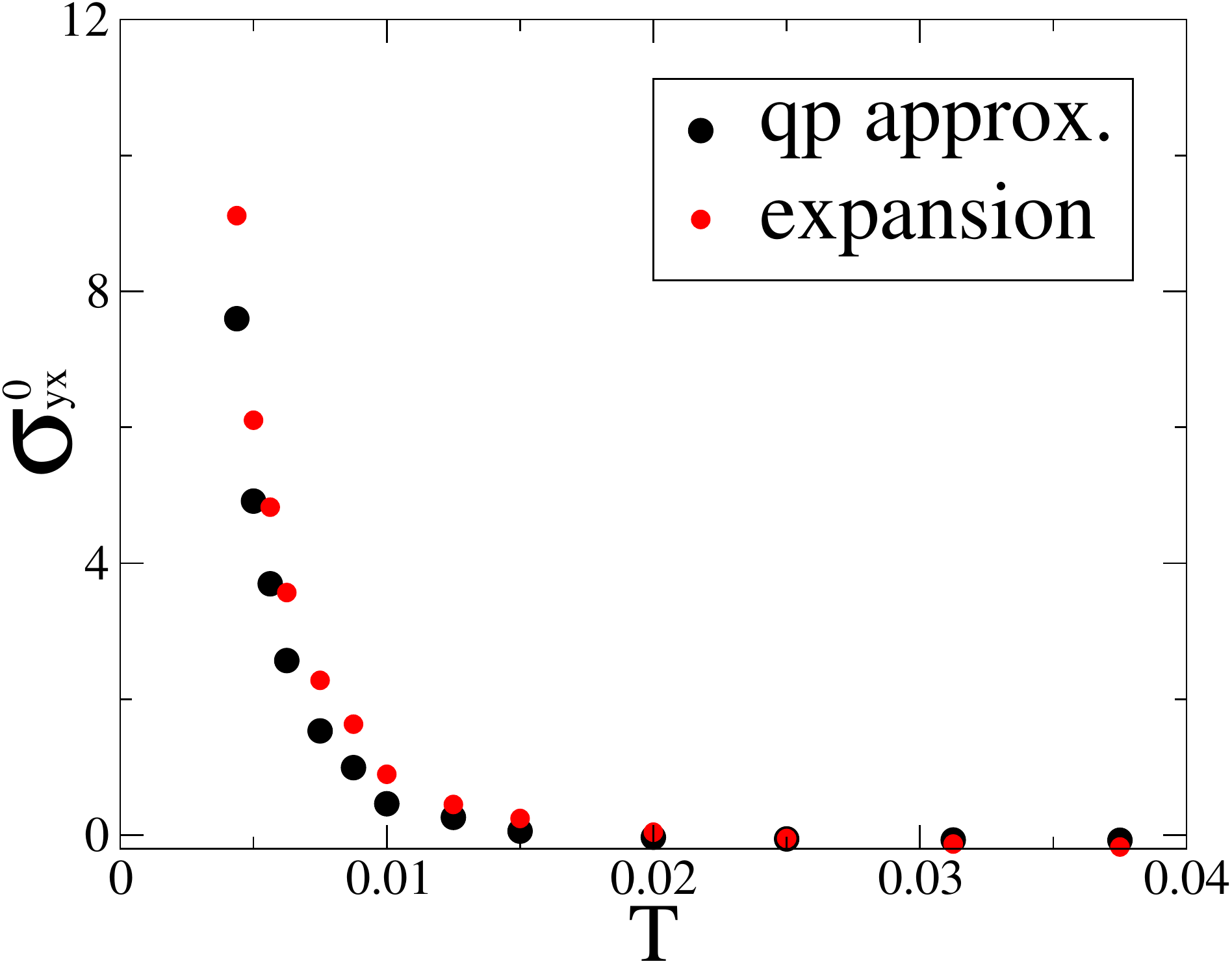} \\
   \includegraphics[width=0.22\textwidth]{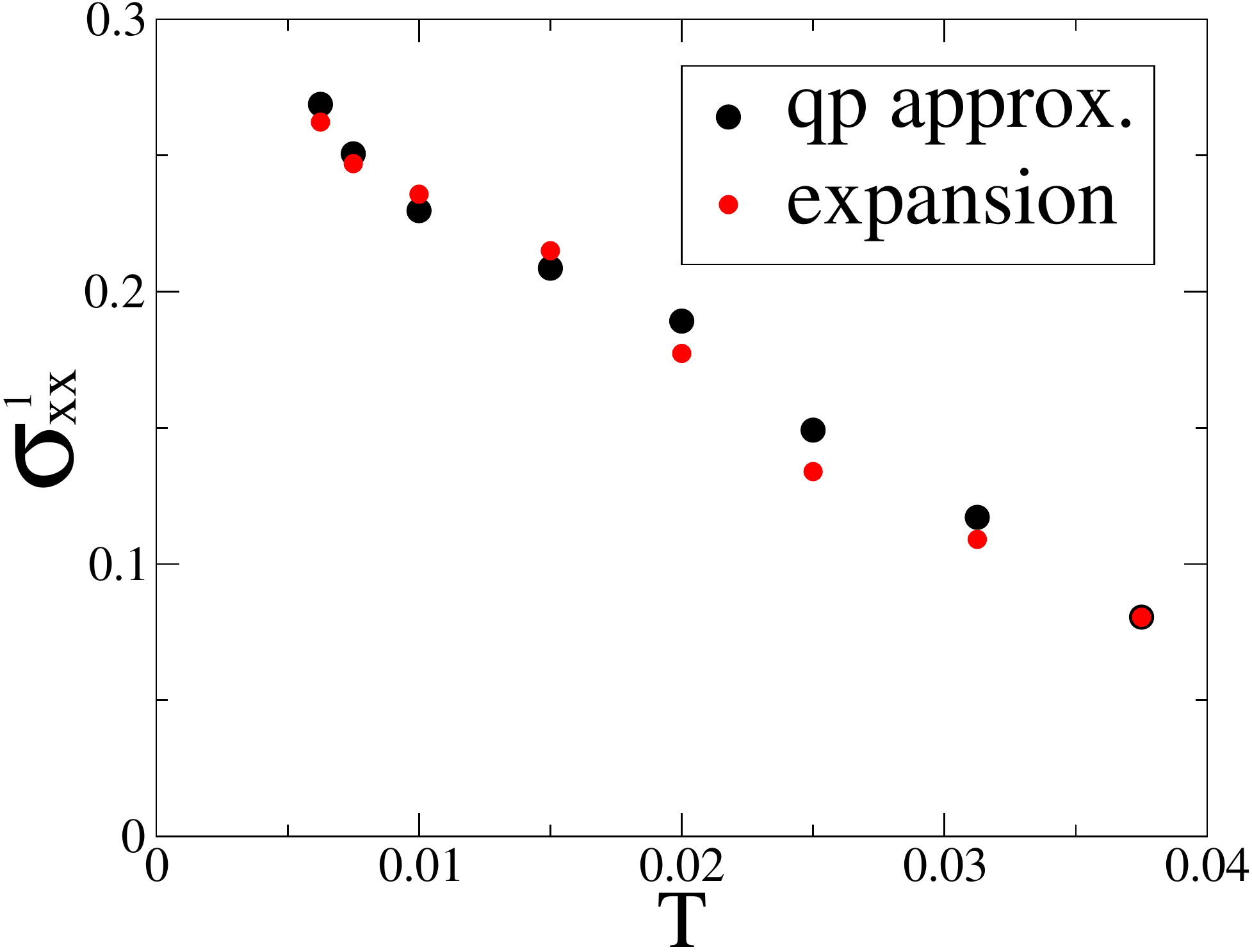} &\includegraphics[width=0.22\textwidth]{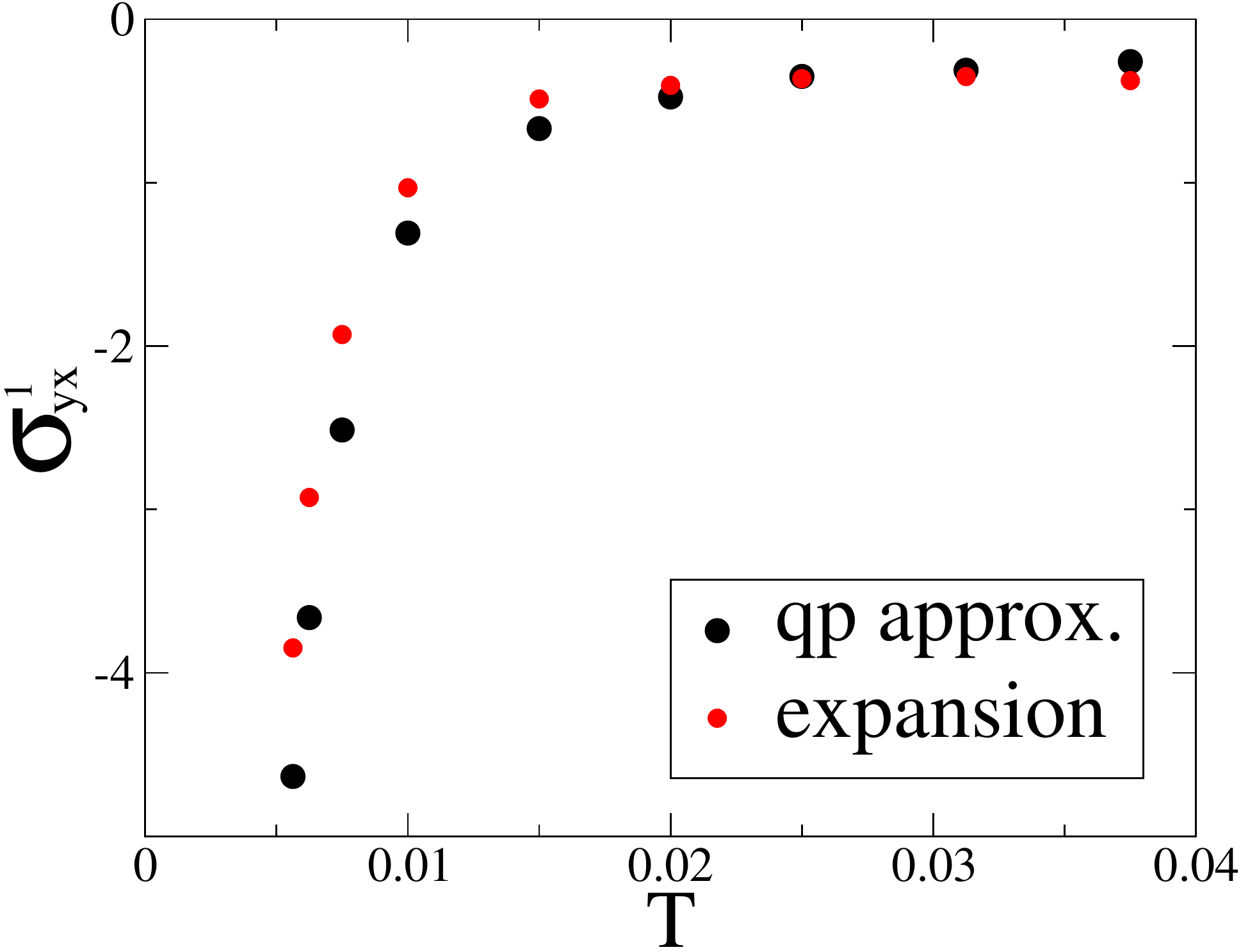}
  \end{tabular}
  \caption{\label{sup:fig:conductivities} Conductivities computed using Eq.~\ref{sup:eq:In} (``qp approx.'') and using the expansion of Eq.~\ref{sup:eq:sigma_xx} ---~\ref{sup:eq:alpha_yx}. }
 \end{figure}

We compute conductivities for $T \lesssim T_{sat}/2$ using Eq.~\ref{sup:eq:In} (``qp approx.'') and using the expansion of Eq.~\ref{sup:eq:sigma_xx} --- Eq.~\ref{sup:eq:alpha_yx}. The results are compared in Fig.~\ref{sup:fig:conductivities}. The agreement between the two sets of data is evident. Therefore Eq.~\ref{sup:eq:sigma_xx} --- Eq.~\ref{sup:eq:alpha_yx} provide a convenient way to decompose the transport properties. In our calculation, $\sigma_{xx}^0$ is dominated by the first term of the expansion. Hence the resistivity is determined by $\xxzero$ and the quasiparticle scattering rate at the Fermi level $\Gamma^*(0)$ (equivalent to $\Gamma_{\bfk_{F}}$ defined in the main text). Similar observation can be drawn for $\sigma_{yx}^0$. In $\sigma_{xx}^{1}$, the $g_3$ term is negligible but the first and the $g_1$ term have opposite signs and both change sign below $T_{sat}/2$. This is due to the different temperature dependence of the asymmetry of the quasiparticle band structure and scattering rate and 
they together determine $\sigma_{xx}^1$ and thermopower. $\sigma_{yx}^{1}$ is dominated by the first term because $\yxone$ is close to its singularity at half-filling. 
\bibliography{reference.bib}